\documentclass{tPHM2e}

\usepackage{graphicx}
\newcommand{\BEQ}{\begin{equation}}
\newcommand{\EEQ}{\end{equation}}
\newcommand{\BEA}{\begin{eqnarray}}
\newcommand{\EEA}{\end{eqnarray}}

\begin{document}
\title{Bond diluted Levy spin-glass model and a new finite size
scaling method to determine a phase transition}
\date{\today}

\author{L. Leuzzi$^{\rm a,b}$$^{\ast}$\thanks{$^\ast$Corresponding
author. Email: luca.leuzzi@cnr.it}, G. Parisi$^{\rm a,b,c}$,
F. Ricci-Tersenghi$^{\rm a,b,c}$ and J.J. Ruiz-Lorenzo$^{\rm d}$
\\
\vspace{6pt} $^{\rm a}${\em{IPCF-CNR, UOS Roma, Piazzale A. Moro 2,
00185, Rome, Italy. }}
\\ $^{\rm b}${\em{Department of Physics,
University ``Sapienza'', Piazzale A. Moro 2, 00185, Rome, Italy.}}\\
$^{\rm c}${\em{INFN, Piazzale A. Moro 2, 00185, Rome, Italy.}}\\
$^{\rm d}${\em{Departamento de F\'{\i}sica, Univ. Extremadura,
Badajoz, E-06071 and BIFI, Spain.}}  }

\maketitle

\begin{abstract}
A spin-glass transition occurs both in and out of the limit of
validity of mean-field theory on a diluted one dimensional chain of
Ising spins where exchange bonds occur with a probability decaying as
the inverse power of the distance.  Varying the power in this
long-range model corresponds, in a one-to-one relationship, to change
the dimension in spin-glass short-range models.  Using different
finite size scaling methods evidence for a spin-glass transition is
found also for systems whose equivalent dimension is below the upper
critical dimension at zero magnetic field. The application of a new
method is discussed, that can be exported to systems
in a magnetic field.
\end{abstract}

\section{Introduction}
Long-range (LR) models are such that their lower critical dimension is
lower than the one of the corresponding short-range (SR) model. In
particular, one can have a phase transition even in one dimensional
systems, provided the range of interaction is large enough.  One
dimensional models with power-law decaying interactions actually allow
to explore both LR and SR regimes by changing the power and enable to
compare the ordered phase in and out of the range of validity of the
mean field approximation.
This is very useful for spin glass models that are known to have a
rather complex states structure in the low-temperature phase in
mean-field theory \cite{Parisi80,MPV}. Whether this
structure exists in finite-dimensional models with short-range
interactions is, though, still a matter of debate.  Theories
alternative to the mean-field one have been proposed
\cite{Fisher86,Krzakala00}, but SR systems are very tough to study
analytically.  Numerical simulations have been, thus, extensively
employed, though with no conclusive indication on the nature of the
spin-glass (SG) phase in finite dimension nor on the existence of a
thermodynamic SG phase in presence of an external magnetic field.

A one-dimensional spin-glass model with power-law decaying
interactions
\cite{vEnter83,Kotliar85,Campanino87,Leuzzi99,Katzgraber03} makes
probes on larger linear sizes possible.  Moreover, if bonds are diluted
\cite{Leuzzi08,Leuzzi09,KYultimo,Larson10} the run time in numerical
simulations grows simply as the size $L$ of the system, rather than
proportionally to $L^2$, as in fully connected systems.  This is a
fundamental issue because finite volume effects are strong in these
models, so much that the very existence of the transition can be
rather difficult to  establish with canonical methods, e.g., when
 the interaction has a rapid decay and/or an external  magnetic field is
applied.

\section{The 1D Levy Spin-Glass Model}

The model investigated is a one dimensional chain of $L$ Ising spins
($\sigma_i=\pm 1$) and Hamiltonian \BEQ {\cal
H}=-\sum_{ij}J_{ij}\sigma_i\sigma_j
\label{f:ham}
\EEQ 
The quenched random couplings $J_{ij}$ are independent and identically
distributed random variables taking a non zero value with a
probability decaying with the distance between spins $\sigma_i$ and
$\sigma_j$, $r_{ij}=|i-j| \mod (L/2)$, as
\begin{equation}
\mathbf{P}[J_{ij}\neq 0] \propto r_{ij}^{-\rho}\quad \text{for}\;\; r_{ij}
\gg 1\;.
\label{eq:Jij}
\end{equation}
Non-zero couplings take value $\pm 1$ with equal probability.  We use
periodic boundary conditions and a $z=6$ average coordination
number. Links are generated by repeating $zL/2$ times the following
process: choose randomly 2 spins at distance $r$ with probability
$r^{-\rho}/\sum_{i=1}^{L/2}i^{-\rho}$; if they are already connected,
repeat the process, otherwise connect them.
\\
\indent
As the power $\rho$ varies this model is known to display different
statistical mechanics behaviors
\cite{vEnter83,Kotliar85,Campanino87}. For the diluted case
\cite{Leuzzi08} they are reported in Tab.~\ref{tab:rho}.
Equivalence between one-dimensional systems with power-law decaying
interactions and $D$-dimensional systems with short-range (nearest
neighbor) interactions can be approximately written as
\cite{KYultimo}:
\BEQ \rho-1=\frac{2-\eta}{D} \EEQ 
where $\eta$ is the critical exponent of the space correlation
function for the short-range model.  The relationship is exact at the
mean-field threshold $\rho=4/3$ ($D=6$, $\eta = 0$) and approximated
below.  The analogue of a 3D spin-glass in zero magnetic field (with
$-0.384(9)<\eta<-0.337(15)$
\cite{Marinari98,Ballesteros00,Jorg06,Katzgraber06,Hasenbuch08}) would
then be a system with $\rho\simeq 1.8$. We will focus on this value in
the present manuscript to introduce and test our method.  Other model
cases have been considered in Refs. \cite{Leuzzi08,Leuzzi09}.
\begin{table}[h!]
\centering
\begin{tabular}{||c|c|c||}
\hline
$\rho$ & $D(\rho)$ &transition type\\
\hline
$\leq 1$ &   $\infty$  &Bethe lattice like\\
$]1:4/3]$ & $[6:\infty[$ & $2^{\rm nd}$ order, MF\\
$]4/3:2]$ & $[2.5:6[$ & $2^{\rm nd}$ order, non-MF\\
$2$ & $2.5$ &Kosterlitz-Thouless or $T=0$-like\\
$>2$ & $<2.5$ &none \\
\hline 
\end{tabular}
\vspace{.2 cm}
\protect\caption{From infinite range to short range behavior of the
  SG model defined in Eqs.(\ref{f:ham},\ref{eq:Jij}).}
\label{tab:rho}
\end{table}

\section{Numerical simulations}
We simulate two replicas $\sigma_i^{(1,2)}$ using the parallel
tempering algorithm (PT) \cite{PT}.  The simulated sizes are $L=2^\kappa$,
with $\kappa=6,8,10,12$. The interval between temperatures in the PT
evolution is $\Delta T=0.05$. The number of samples is
$N_J=6400-90000$ depending on the size.  All data used for our
analysis are thermalized. Thermalization has been checked by measuring
all observables on exponentially growing time windows until the last
two points coincide within the statistical error.

\section{Critical point}
\begin{figure}[b!]
\centering
\includegraphics[width=.65\textwidth]{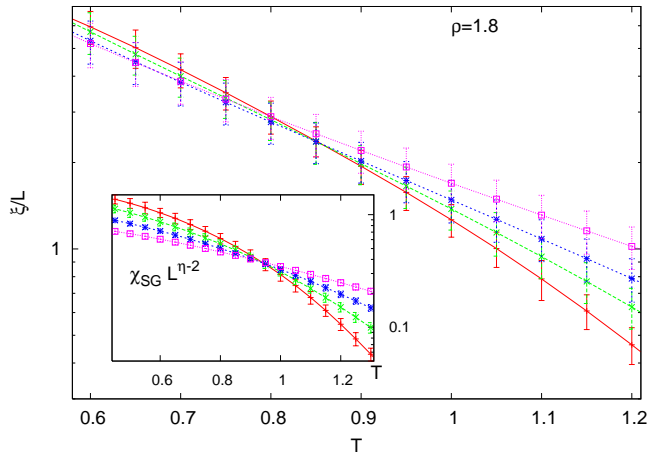}
\caption{FSS function $\xi/L$ vs. $T$ for different simulated sizes at
$\rho=1.8$ for $L=2^{6,8,10,12}$. Inset: $\chi_{\rm SG}L^{\eta-2}$ vs. $T$ for
the same sizes.}
\label{fig:xisuL}
\end{figure}

The key observable to approach the critical behavior in disordered systems is
 the four-spins correlation function: 
\BEQ
C_4(x)=\frac{1}{L}\sum_{i=1}^L{\overline{\langle
\sigma_i^{(1)}\sigma_i^{(2)}\sigma_{i+x}^{(1)}\sigma_{i+x}^{(2)}\rangle}} 
\EEQ 
and its Fourier transform $\tilde C_4(k)$.
In order to determine the critical point,
a correlation length-like observable is usually defined on the 1D lattice as
\cite{Caracciolo93,Katzgraber03}
 \BEQ \xi = \frac{1}{2\sin
k_1/2}\left[\frac{\tilde C_4(0)}{\tilde
C_4(k_1)}-1\right]^{1/(\rho-1)} 
\label{f:xi}
\EEQ
with $k_1=2\pi/L$.  In Fig. \ref{fig:xisuL} we present the $\xi/L$
curves whose crossing point should tend, as $L\to\infty$, to $T_c$. In
the inset we also show the behavior of
 \BEQ \chi_{\rm SG}=L\tilde C_4(0),
\EEQ
 another finite size scaling (FSS) function for the present model (in which
$\eta=\eta_{\mathrm{MF}}=\rho-1$ also for $\rho>\rho_{\mathrm{MF}}$).
Due to the statistical error it is not straightforward to identify
clear crossing points for $\xi/L$. Moreover, in both above mentioned
cases, to extrapolate a clear limit of $T_c$ as $L\to\infty$ with a
FSS interpolating function like $a+bL^{-c}$,
cf. Fig. \ref{fig:Tc_comp}, three degrees of freedom are not enough
and the interpolations are thus just indicative (see the following).

\section{A novel method}
\begin{figure}[t!]
{\includegraphics[width=.48\textwidth]{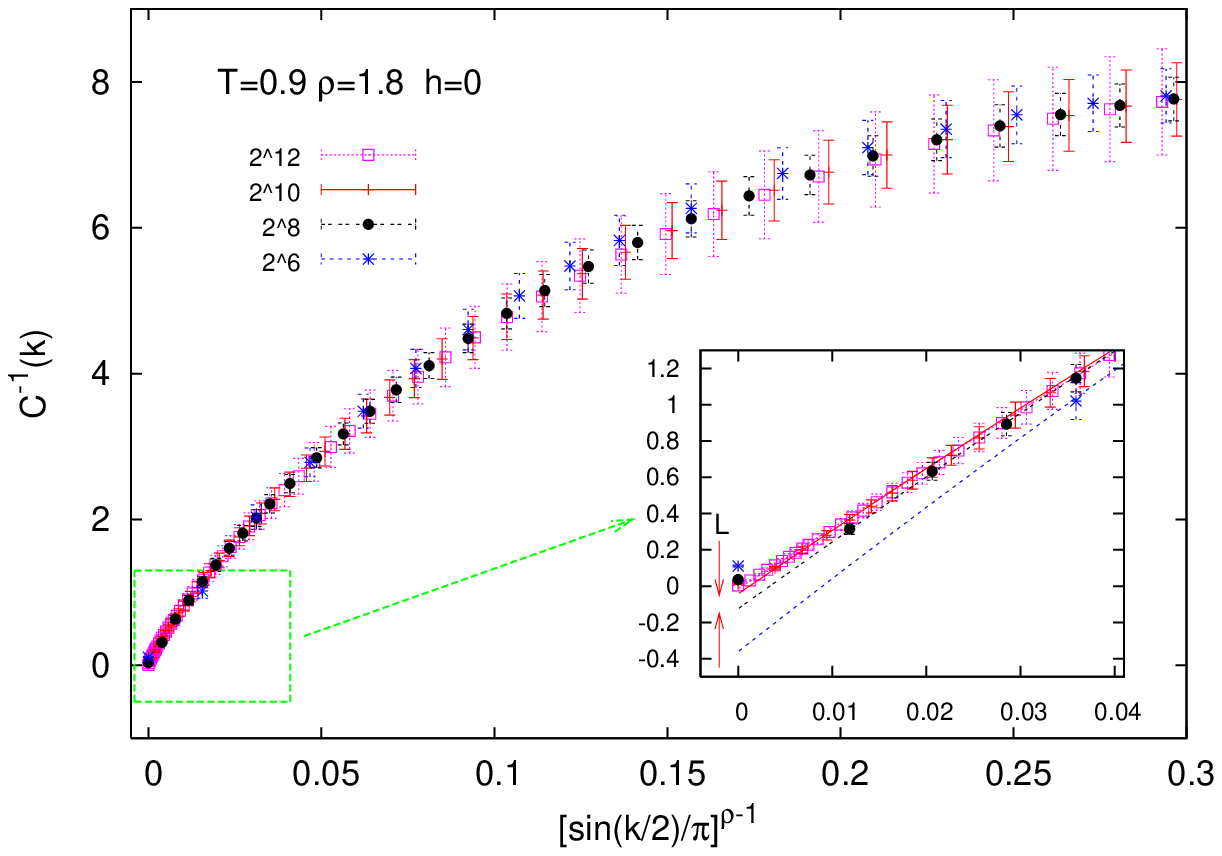}
\includegraphics[width=.48\textwidth]{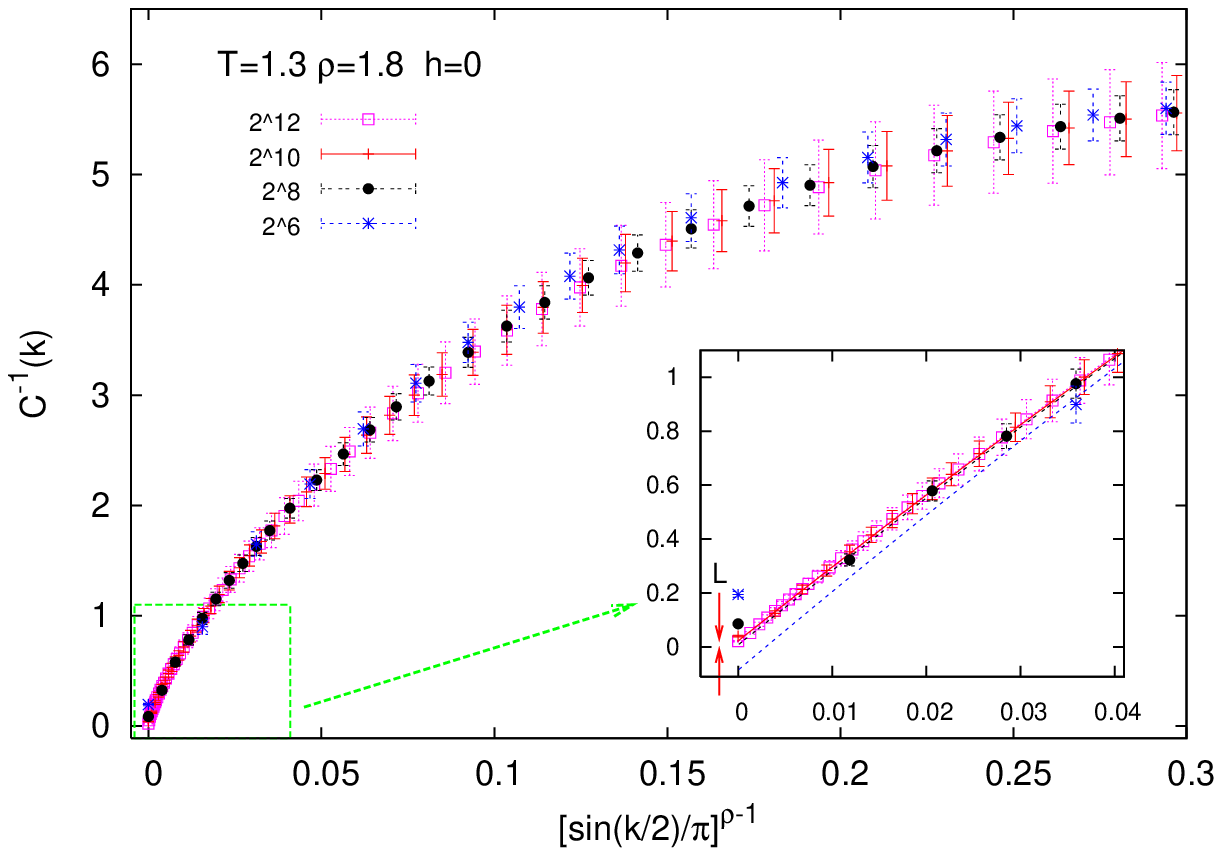}}
\caption{Average $\tilde C^{-1}_4(k)$ vs. $[\sin(k/2)/\pi]^{0.8}$
 for systems of size $L=2^{6,8,10,12}$. Left: $T=0.9$, slightly below the
 critical region. Right: $T=1.3> T_c$. Insets: detail for low $k$
 (points) and comparison of the values in $k=0$ with the $k\to 0$
 limit of the interpolation $A+B [\sin(k/2)/\pi]^{0.8}$ (lines).}
\label{fig:tildeC}
\end{figure}

We can, otherwise, use the whole information provided by the $\tilde
C_4(k)$. In Fig. \ref{fig:tildeC} we plot $1/\tilde C_4(k)$
vs. $[\sin(k/2)/\pi]^{\rho-1}$ for $\rho=1.8$ for all simulated sizes
both at temperature above $T_c$ (right) and at a $T\simeq T_c$ (left).
We observe that finite size effects act in opposite ways on the value of
$\tilde C_4(0)$ (i.e., $\chi_{\rm SG}$) and on the rest of the function
$\tilde C_4(k>0)$ (see insets of Fig.  \ref{fig:tildeC}): while
$1/\tilde C_4(0)$ tends to its thermodynamic limit from above,
$1/\tilde C_4(k>0)$ and its interpolation at $k=0$ tend to the thermodynamic
limit from below. Even though $\tilde C_4(0)$ and $\lim_{k\to 0} \tilde
C_4(k)$ are the same object for $L\to \infty$ their FSS scaling is
qualitatively different.  

\begin{figure}[b!]
\centering
\includegraphics[width=.65\textwidth]{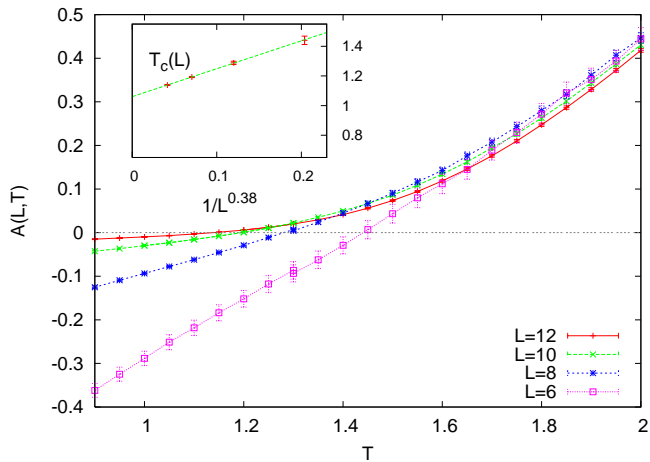}
\caption{Behavior of $A(L,T)$, cf. Eq. (\ref{f:A}), for
$L=2^{6,8,10,12}$. The points at which $A=0$ are finite size estimates
of $T_c$.}
\label{fig:ALT}
\end{figure}

Interpolating $1/\tilde C_4(k)$ for small $k$ at a given size and
temperature as 
\BEQ
F^{\rm fit}(k) = A(L,T)+B(L,T) [\sin(k/2)/\pi]^{0.8}
\EEQ
we can analyze the $L$ and $T$ dependence of 
\BEQ
A(L,T)\equiv  \lim_{k\to 0} \tilde C_4(k)
\label{f:A}
\EEQ and determine the transition from the FSS analysis of the points
at which $A(L,T)=0$, rather than using the FSS of the crossing points
of Eq. (\ref{f:xi}) (or $\tilde C_4(0) L^{\eta-1}$).
 For $\rho=1.8$, the behavior of $A$ in $L$ and $T$ is plotted in
Fig. \ref{fig:ALT}.

\begin{figure}[t!]
\centering
\includegraphics[width=.65\textwidth]{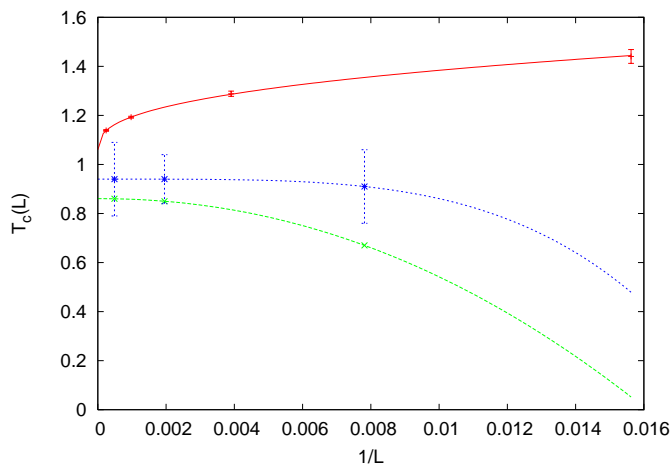}
\caption{Comparison between the finite size estimates of $T_c(L)$ by
means of the canonical ``crossings'' methods (of the FSS functions
$\xi/L$ and $\chi_{\rm SG}L^{\eta-2}$) and of the $A(L,T)=0$ method,
cf. Eq. (\ref{f:A}).  The fit for $T_c$ from $A(L,T)=0$ has a
$\chi^2=0.075$ (full/red curve), while the other two interpolating
curves (dashed/green for $\xi/L$, dotted/blues for $\chi_{\rm
SG}L^{\eta-2}$) are only indicative (not enough d.o.f. to provide an
estimate of statistical errors, no error bars defined on the $T_c(L)$
from $\xi/L$ crossings).  }
\label{fig:Tc_comp}
\end{figure}

This method has the advantage of using high temperature data and one
only needs to simulate systems down to the candidate $T_c$.  As
$A(\infty,T)$ becomes negative, indeed, the functional form of the
propagator in the paramagnetic phase breaks down and this provides
evidence for a phase transition.  In Fig. \ref{fig:Tc_comp} we plot
the finite size values of $T_c$ obtained by this method and we compare
them with the estimates derived from FSS functions $\xi$ and
$\chi_{\rm SG}$. At $\rho=1.8$ we find $T_c = 1.060(7)$ and
$1/\nu=0.38(2)$ with a $\chi^2=0.075$ on the same data of the previous
analysis (same statistics, same thermalization times, same sizes).

We stress that also the definition of $\xi$ as a correlation length in
Eq. (\ref{f:xi}) is valid only in the paramagnetic phase and that
below $T_c$ this  is just a scaling function without physical meaning.
In that approach, though, in order to appreciate crossings of
$\xi_L(T)/L$ curves at different $L$ one has to simulate the system
also at temperatures below $T_c$, where thermalization times
increase, and for at least five different sizes in order to provide
enough points for a well defined FSS interpolation.

\section{Correlation length estimate}

\begin{figure}[t!]
\centering
\includegraphics[width=.65\textwidth]{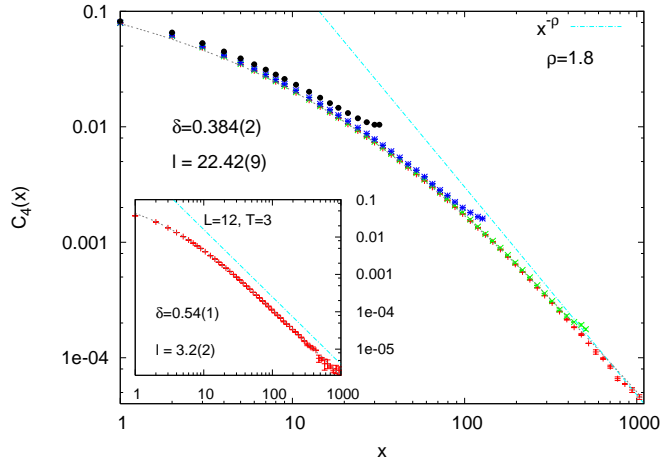}
\caption{Four spins correlation function at high temperature
($T=2\simeq 1.9 T_c$) in log-log scale for all simulated sizes
$L=2^{6,8,10,12}$. Dotted lines: interpolation by means of
Eq. (\ref{f:fit_C4}).  Dashed-dotted line: $x^{-\rho}$.
No finite size effects are present, apart from the
values at $x\simeq L/2$, and a crossover from a power law with
exponent $\alpha\simeq 0.2$ to a power law with $\rho=1.8$ can be
identified for $L>10$. Inset: $C_4(x)$ for $L=2^{12}$ at $T=3$ compared to
$x^{-1.8}$. }
\label{fig:C4_x_hT}

\end{figure}

We are interested in 
characterizing the above mentioned critical
behavior by means of a growing correlation length, as it happens in
ordinary continuous phase transitions in systems without quenched
disorder. Also for this analysis our starting point is the four-spins
correlation function, this time in position space.  As one can notice
from Fig. \ref{fig:C4_x_hT}, we can identify two different decays of
the correlation at a given $T>T_c$, if we are able to study long
enough Levy-glass chains.  We observe a slower power-law decay as
$x\ll L$, $x^{-\alpha}$, and a faster decay as $x\sim L/2$.
Contrarily to what happens in short-range models at $D\geq 2.5$
\cite{Marinari98,Contucci07}, this second decay is also power-law
(with power equal to $\rho$) because the interaction correlation
decays - by construction - as $x^{-\rho}$ and, therefore, the $C_4(x)$
cannot decay any faster.

 \begin{figure}[t!]
\centering
\includegraphics[width=.65\textwidth]{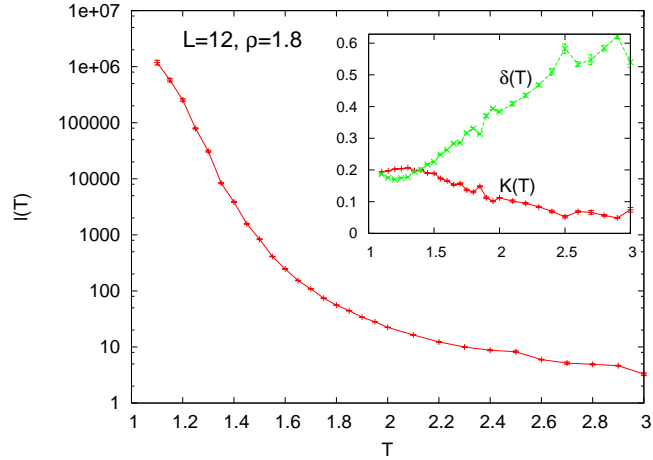}
\caption{Correlation length parameter $\ell$ of interpolating function
$C_4^{\rm fit}(x)$ vs. $T$ in log scale for 
$L=2^{12}$. Inset: $T$ behavior of fir parameters $K$ and $\delta$.}
\label{fig:ell}
\end{figure}

From the definition of $\eta$ as exponent of the anomalous decay of
the correlation function as $x^{-d+2-\eta}$, with $\eta=3-\rho$, we
obtain the relation $\alpha+\rho=d+1=2$ \cite{footnote}.  We can, thus,
think about interpolating the whole $C_4(x)$ behavior above the
critical point as \BEQ C_4^{\rm fit}(x) \equiv K x^{-\alpha}\left[
1+\left(\frac{x}{\ell}\right)^{2\delta(\rho-\alpha)}
\right]^{-1/\delta}
\label{f:fit_C4}
\EEQ
with $\rho=1.8$ and $\alpha = 2 - \rho = 0.2$.

From this interpolation we can look at the temperature behavior of the
length-like parameter $\ell$, that is an estimate of the correlation
length of the system, as far as the second power-law decay is
observable ($T>T_c$). In Fig. \ref{fig:ell} we display the behavior of
$\ell(T)$ (and $K$ and $\delta$ in the inset) as $T$ is lowered down
to the estimated $T_c$. Fits with Eq. (\ref{f:fit_C4}) are reasonable
down to $T=1.1$.

\begin{figure}[t!]
\centering
\includegraphics[width=.65\textwidth]{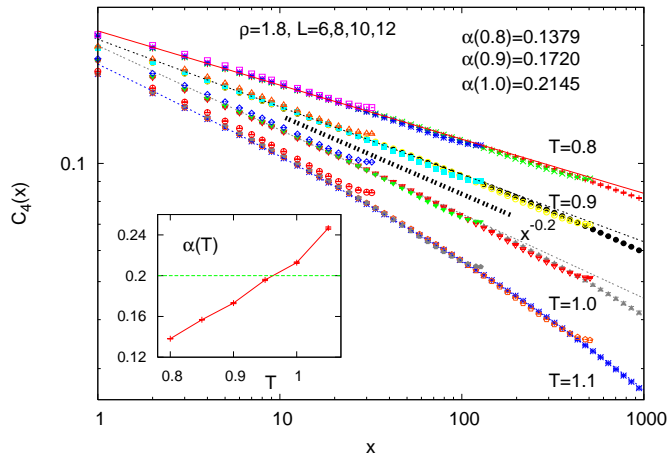}
\caption{Four spins correlation function across the critical region
($T=0.8,0.9,1,1.1$) in log-log scale for all simulated sizes
$L=2^{6,8,10,12}$ and relative interpolating functions.  At all $T$,
finite size effects are seemingly small. Dotted lines: interpolation
by means of Eq. (\ref{f:fit_C4}) at $T=1.1$; for lower $T$ the fit
function is $A x^{-\alpha}$ is used and no crossover to a power law
with $\rho=1.8$ is observed for any $L$, in agreement with $\ell\gg
L$. The $T$-dependence of the value of $\alpha$ is shown in the
inset.}
\label{fig:C4_x_Tc}
\end{figure}

As temperature approaches the critical value from above, the simulated
systems are too small to appreciate the existence of a crossover
length $\ell \gg L$ and, thus, the $C_4(x)$ decay appears as a single
power law. The same behavior remains below $T_c$, as shown in
Fig. \ref{fig:C4_x_Tc}. This is incompatible with the onset of a
plateau at any $x$ whereas it is consistent with the clustering
properties of the mean-field theory for spin-glasses.

\section{Conclusions}
We have introduced a new FSS method to determine, by numerical
simulations, the existence of a critical point in finite dimensional
systems. Such method employs high temperature data.  Thus, it requires
lower thermalization times and disordered sample statistics with
respect to canonical methods. It works well if one has a sufficient
amount of points of the four-spins correlation function at low
wavelength numbers $k$, that is, if the interaction range is not too
broad (i.e., preferably out of mean-field) and linear size is long
enough.  
\\
\indent
We have tested the method in the case of a bond-diluted 1D
Levy spin-glass with a power-law decaying interaction outside the
limit of validity of mean-field approximation and equivalent to a
3D nearest-neighbor interacting system on a cubic lattice.  We have
compared the results with those obtained by canonical FSS analysis on
the same set of data and shown that they are compatible,
cf. Fig. \ref{fig:Tc_comp}. 
\\
\indent
In position space, at $T>T_c$ we have
identified a crossover in the (power-law) relaxation decay of the
four-spins correlation function from slow to fast (i.e., $x^{-\rho}$)
relaxation, cf. Eq. (\ref{f:fit_C4}). The crossover takes place at a
correlation length that grows exponentially as $T$ decreases. At
$T<T_c$ the decay is well interpolated by a simple power-law,
providing no evidence for the plateau predicted by droplet
\cite{Fisher86} and TNT theories \cite{Krzakala00}.

\section{Acknowledgments}
This work was partially supported by the Ministerio de Ciencia y
Tecnolog\'{\i}a (Spain) through Grant No. FIS2007-60977, by the Junta de
Extremadura (Spain) through Grant No. GRU09038 (partially founded by
FEDER).


\begin{thebibliography}{99}
\bibitem{footnote} This can be
seen, e.g., in Ref. \cite{Leuzzi08} for $\rho=1.5$. In the inset
of figure 6 $\alpha$ was erroneously misplaced with $\delta$.
\end{thebibliography}
\end{document}